\documentstyle[preprint,aps,eqsecnum]{revtex}
\begin{document}
\draft
\title{
\begin{flushright}
{\normalsize ~~~%IP-ASTP-16-95
}
\end{flushright}
\Large\bf
$B \rightarrow \pi l {\nu}$ Form Factors Calculated
on the Light-Front
}
\author{
{\bf Chi-Yee Cheung},$^a$
%\footnote{e-mail address: phcheung@ccvax.sinica.edu.tw}
{\bf Chien-Wen Hwang},$^{a,b}$
and {\bf Wei-Min Zhang}~$^a$ \\
%\footnote{e-mail address: wzhang@phys.sinica.edu.tw}
{\it $^a$Institute of Physics, Academia Sinica, Taipei 11529, Taiwan}\\
{\it $^b$Department of Physics, National Taiwan University, Taipei 10764,
Taiwan}
}
%
%\date{\today}
%
\maketitle
\begin{abstract}

A consistent treatment of $B\rightarrow \pi l \nu$ decay
is given on the light-front.
The $B$ to $\pi$ transition form factors are calculated
in the entire physical range of momentum transfer for the first time.
The valence-quark contribution
is obtained using relativistic light-front wave functions.
Higher quark-antiquark Fock-state of
the $B$-meson bound state is represented
effectively by the $|B^*\pi\rangle$ configuration, and its effect
is calculated in the chiral perturbation theory.
Wave function renormalization is taken into account consistently.
The $|B^*\pi\rangle$ contribution dominates near the zero-recoil
point ($q^2\simeq 25$ GeV$^2$),
and decreases rapidly as the recoil momentum increases.
We find that the calculated form factor $f_+(q^2)$ follows approximately
a dipole $q^2$-dependence in the entire range of momentum transfer.

\vskip 0.25 true cm
PACS numbers: 13.20, 14.40.J
\end{abstract}
\newpage
\baselineskip .29in
%%%%%%%%%%%%%%%%%%%%%%%%%%%%%%%%%%%%%%%%%%%%%%%%%%%%%%%%%%%%%%%%%%%%%%%%%%%%%%%

\section{Introduction}

The study of exclusive semileptonic decays of heavy mesons
has attracted much interest in recent years.
Semileptonic decays of heavy mesons to heavy mesons,
such as $B\rightarrow D(D^*) l \nu$,
provide an ideal testing ground for heavy-quark symmetry
and heavy-quark effective theory.
By comparison, weak decays of heavy mesons to light mesons
are much more complicated theoretically,
since in general there exists no symmetry principle for guidance.
Nevertheless, it is essential to understand the
reaction mechanisms of these decay modes, because they are
the main sources of information on the CKM mixing matrix elements
between heavy and light quarks.  In particular, the study of
the $B\rightarrow\pi l \nu$ decay
is important for the determination of matrix
element $V_{ub}$ whose value is only poorly known\cite{PDG}.

Recently, the $B \rightarrow \pi l {\nu}$ decay
has been investigated by many groups
\cite{BSW,ISGW,Odon,Faus,Ball,Che,Bel1,Col,Ali,HYC,Neu,Wise,BD1,Wol,Casa,LY}
using various different approaches, such as quark models,
QCD sum rules, heavy-quark symmetry, perturbative QCD, and so on.
In most studies,
transition form factors are calculated only at one
kinematic point, $q^2 = (P_B - P_{\pi})^2 = 0$,
so that extra assumptions are needed to extrapolate the
form factors to cover the entire range of momentum transfer.
In \cite{Wise,BD1,Wol,Casa} chiral perturbation is employed,
so that the results
are valid for soft pion emission only.
In \cite{LY} perturbative QCD is applied, in conjunction
with hadronic wave functions obtained from QCD sum rule,
to calculate the decay amplitudes; therefore the result is valid
only when the final pion is energetic.

In this study, we first calculate the $B\rightarrow\pi l \nu$
decay form factors using relativistic
light-front hadronic wave functions [Fig. (1a)].  The parameters in
these wave functions are determined from other informations,
and Melosh transformation is used to construct meson
states of definite spins.
Although light-front wave functions
and quark model in the infinite momentum frame
have been used in the past
to study $B \rightarrow \pi l {\nu}$ decay
and other heavy-light transitions\cite{BSW,Odon,Jaus},
the decay form factors were only calculated for $q^2=0$.
In this work, we directly evaluate for the first time
the form factors in the entire kinematic
region, so that additional extrapolation assumptions are no longer required.
Secondly, we note that the $B\rightarrow\pi$ transition involves time-like
momentum transfers.  That means $q^+\geq 0$ in the light-front coordinate,
so that one must also consider the effects of the so-called
nondiagonal light-front 
diagrams (or Z-graphs)\cite{Jaus,Brodsky,Sawicki,Dubin},
as depicted in Fig. (1b).
These contributions are generated by
the quark-antiquark ($q\bar q$) excitation or higher Fock-states
in the hadronic bound states.
We shall effectively represent the $q\bar q$-
configuration of the $B$-meson
by the mesonic $|B^*\pi\rangle$ state,
giving rise to the $B^*$-pole contribution shown in Fig. (1c).
It was first noted by Isgur and Wise\cite{IW} that
the $B^*$-pole effect is important in the zero-recoil region.
Previous investigations either simply added this
effect to the valence-quark contribution\cite{IW,BD2},
or totally ignored it.
In our more unified approach,
the $B^*$-pole contribution arises from the $|B^*\pi\rangle$-component
of the $B$-meson bound state.
The mixing of different Fock-state configurations naturally
requires a consistent wave function renormalization
which has not been mentioned in all the previous works.

This paper is organised as follows.  In Section II, the basic
theoretical formalism is given.  In Section III, numerical
results are present and discussed, and finally a summary
is given in Section IV.

%%%%%%%%%%%%%%%%%%%%%%%%%%%%%%%%%%%%%%%%%%%%%%%%%%%%%%%%%%%%%%%%%%%%%%%%%%%%%%%

\section{General Formalism}

The weak current that is responsible for $b$-quark decay is given
by
\begin{equation}
        J^\mu = \bar q\gamma^\mu (1-\gamma_5) b,
\end{equation}
where $q$ stands for a light quark.
For $B\rightarrow \pi l \nu$, only the vector current contributes,
and our main task is to evaluate the hadronic matrix element,
$M^\mu_{B\pi} = \langle\pi(P_\pi)|J^\mu|B(P_B)\rangle$,
which can be parametrized as
\begin{eqnarray}
M^\mu_{B\pi} &=& \langle\pi(P_\pi)|J^\mu|B(P_B)\rangle\nonumber\\
             &=& f_+(q^2) (P_B + P_{\pi})^{\mu} + f_-(q^2)
                (P_B-P_{\pi})^{\mu}. \label{mele}
\end{eqnarray}

In the previous calculations of hadronic matrix elements in
the infinite momentum frame or with light-front wave functions,
one usually set $q^+ = P_B^+ - P_{\pi}^+ =0$.  This leads to
$q^2 = -q_\bot^2$, implying a space-like momentum transfer.
However, momentum transfers in real
decay processes are always time-like.  Hence
matrix elements calculated with $q^+ = 0$
is relevant only at
the maximum-recoil point with $q^2=0$, and one needs
an extrapolation ansatz to extend the result to other
physical momentum transfers.
A direct calculation of the form factors for the whole
momentum transfer range has not been performed.
In this work, we work in a frame where the ``$\bot$"-
components of $\vec P_B$ and $\vec P_\pi$ vanish,
so that $q^2=q^+q^-\ge 0$, and we can evaluate the form factors in
the entire physical range of momentum transfer.
As mentioned earlier, in a frame with $q^+>0$,
there are two distinct contributions to the matrix element.
Apart from the usual valence contribution [Fig. (1a)]
which is calculated with relativistic light-front wave functions,
one must also include the nondiagonal light-front diagram
as depicted in Fig. (1b).
Here, such non-valence effects are taken into account effectively
by the $B^*$-pole contribution shown in Fig. (1c).
Simple perturbation theory in the light-front approach gives
\cite{Zhang}
\begin{equation}
        |B(P_B)\rangle = \sqrt{Z_2} \Bigg\{ |B_0 (P_B)\rangle
                +  \int [d^3 k] [d^3 q]
                { \langle B^*(q) \pi(k) | H_I | B_0(P_B)
                \rangle \over P_B^- - q^- - k^- }
                |B^* (q) \pi (k) \rangle \Bigg\}, \label{wf}
\end{equation}
where
$|B_0\rangle$ represents the valence configuration
described by a light-front bound-state wave function,
while $|B^*~\pi\rangle$ is the most important higher-Fock-state
configuration, as will be explained later;
$H_I$ is the interaction Hamiltonian for the $BB^*\pi$-vertex
obtainable from chiral perturbation theory\cite{Wise,BD1,Yan},
\begin{eqnarray}
      [d^3k] &\equiv& {dk^+ d^2k_\bot \over 2(2\pi)^3 k^+}, \nonumber\\
      \langle P'|P\rangle&=&2(2\pi)^3 P^+\delta(P'^+-P^+)
      \delta^2(P'_\bot-P_\bot),
\end{eqnarray}
and $\sqrt{Z_2}$ is the wave function renormalization constant.
All particles in Eq. (\ref{wf})
are on the mass-shells so that $P_B^- = {P_{B\bot}^2 + M_B^2
\over P_B^+}$, $q^- = {q_{\bot}^2 + M_{B^*}^2 \over q^+}$,
and $k^- = {k_\bot^2 + m_{\pi}^2 \over k^+}$.

From Eqs. (\ref{mele}) and (\ref{wf}), we have formally
\begin{equation}
        M^\mu_{B\pi} = \sqrt{Z_2} \Bigg\{ \Big(f_+^v(q^2) + f_+^{B^*} (q^2)
                \Big) (P_B + P_{\pi})^{\mu} + \Big(f_-^v(q^2) + f_-^{B^*}
                (q^2) \Big) (P_B-P_{\pi})^{\mu} \Bigg\},
\end{equation}
where $f_+^v, f_-^v$ represent the valence-configuration contributions,
and $f_+^{B^*}, f_-^{B^*}$ the $B^*$-pole contributions.
The resultant $B \rightarrow \pi l {\nu}$ decay form
factors are therefore given by
\begin{eqnarray}
        f_+(q^2) &=& \sqrt{Z_2}~\Big[f_+^v(q^2) + f_+^{B^*} (q^2)\Big],
        \nonumber\\
        f_-(q^2) &=& \sqrt{Z_2}~\Big[f_-^v(q^2) + f_-^{B^*} (q^2)\Big].
\end{eqnarray}
Most of the previous investigations only calculated $f_\pm^v$ at
$q^2=0$, some also included the contribution of $f_\pm^{B^*}$, but none
has taken into account the effect of $\sqrt{Z_2}$.

In the following, we shall first calculate the valence contribution
using light-front bound state wave functions.
Subsequently, the $B^*$-pole contribution,
as well as the effect of wave function renormalization, will be
discussed in more details.

\subsection{Valence Configuration Contribution}

A meson bound state consisting of a quark $q_1$ and
an anti-quark $\bar q_2$ with total momentum $P$
and spin $S$ can be written as
\begin{eqnarray}
        |M(P, S, S_z)\rangle
                =\int &&\{d^3p_1\}\{d^3p_2\} ~2(2\pi)^3 \delta^3(\tilde
                P-\tilde p_1-\tilde p_2)~\nonumber\\
        &&\times \sum_{\lambda_1,\lambda_2}
                \Psi^{SS_z}(\tilde p_1,\tilde p_2,\lambda_1,\lambda_2)~
                |q_1(p_1,\lambda_1) \bar q_2(p_2,\lambda_2)\rangle,
\end{eqnarray}
where $p_1$ and $p_2$ are the on-mass-shell light-front momenta,
\begin{equation}
        \tilde p=(p^+, p_\bot)~, \quad p_\bot = (p^1, p^2)~,
                \quad p^- = {m^2+p_\bot^2\over p^+},
\end{equation}
and
\begin{eqnarray}
        &&\{d^3p\} \equiv {dp^+d^2p_\bot\over 2(2\pi)^3} \nonumber \\
        &&|q(p_1,\lambda_1)\bar q(p_2,\lambda_2)\rangle
        = b^\dagger_{\lambda_1}(p_1)d^\dagger_{\lambda_2}(p_2)|0\rangle,\\
        &&\{b_{\lambda'}(p'),b_{\lambda}^\dagger(p)\} =
        \{d_{\lambda'}(p'),d_{\lambda}^\dagger(p)\} =
        2(2\pi)^3~\delta^3(\tilde p'-\tilde p)~\delta_{\lambda'\lambda}.
                \nonumber
\end{eqnarray}
In terms of the light-front relative momentum
variables $(x, k_\bot)$ defined by
\begin{eqnarray}
        && p^+_1=x_1 P^+, \quad p^+_2=x_2 P^+, \quad x_1+x_2=1, \nonumber \\
        && p_{1\bot}=x_1 P_\bot+k_\bot, \quad p_{2\bot}=x_2
                P_\bot-k_\bot,
\end{eqnarray}
the momentum-space wave-function $\Psi^{SS_z}$
can be expressed as
\begin{equation}
        \Psi^{SS_z}(\tilde p_1,\tilde p_2,\lambda_1,\lambda_2)
                = R^{SS_z}_{\lambda_1\lambda_2}(x,k_\bot)~ \phi(x, k_\bot),
\end{equation}
where $\phi(x,k_\bot)$ describes the momentum distribution of the
constituents in the bound state, and $R^{SS_z}_{\lambda_1\lambda_2}$
constructs a state of definite spin ($S,S_z$) out of light-front
helicity ($\lambda_1,\lambda_2$) eigenstates.  Explicitly,
\begin{equation}
        R^{SS_z}_{\lambda_1 \lambda_2}(x,k_\bot)
                =\sum_{s_1,s_2} \langle \lambda_1|
                {\cal R}_M^\dagger(1-x,k_\bot, m_1)|s_1\rangle
                \langle \lambda_2|{\cal R}_M^\dagger(x,-k_\bot, m_2)
                |s_2\rangle
                \langle {1\over2}s_1
                {1\over2}s_2|SS_z\rangle,
\end{equation}
where $|s_i\rangle$ are the usual Pauli spinor,
and ${\cal R}_M$ is the Melosh transformation operator:
\begin{equation}
        {\cal R}_M (x,k_\bot,m_i) =
                {m_i+x_iM_0+i\vec \sigma\cdot\vec k_\bot \times \vec n
                \over \sqrt{(m_i+x_i M_0)^2 + k_\bot^2}},
\end{equation}
with $\vec n = (0,0,1)$, a unit vector in the $z$-direction, and
\begin{equation}
        M_0^2={ m_1^2+k_\bot^2\over x_1}+{ m_2^2+k_\bot^2\over x_2}.
\end{equation}
It is possible to rewrite the transformation matrix
$R^{SS_z}_{\lambda_1\lambda_2}$ in a covariant form\cite{Jaus},
which is useful in practical calculations:
\begin{equation}
        R^{SS_z}_{\lambda_1\lambda_2}(x,k_\bot)
                ={\sqrt{p_1^+p_2^+}\over \sqrt{2} ~{\tilde M_0}}
        ~\bar u(p_1,s_1)\Gamma v(p_2,s_2), \label{covariant}
\end{equation}
where
\begin{eqnarray}
        &&{\tilde M_0} \equiv \sqrt{M_0^2-(m_1-m_2)^2}, \nonumber\\
        &&\bar u(p,s) u(p,s')={2m\over p^+} \delta_{s,s'},
                \quad \sum_s u(p,s) \bar u(p,s)
                ={\not{\! p}+m\over p^+},\nonumber\\
        &&\bar v(p,s) v(p,s')= - {2m\over p^+}\delta_{s,s'},
                \quad \sum_s v(p,s) \bar v(p,s)
                ={\not{\! p}-m\over p^+},\\
        &&\Gamma=\gamma_5 \qquad ({\rm pseudoscalar}, S=0), \nonumber\\
        &&\Gamma=-\not{\! \varepsilon}(S_z)+{\varepsilon\cdot(p_1-p_2)
                \over M_0+m_1+m_2} \qquad ({\rm vector}, S=1), \nonumber
\end{eqnarray}
with
\begin{eqnarray}
        &&\varepsilon^\mu(\pm 1) =
                \left[{2\over P^+} \vec \varepsilon_\bot (\pm 1) \cdot
                \vec P_\bot,0,\vec \varepsilon_\bot (\pm 1)\right],
                \quad \vec \varepsilon_\bot
                (\pm 1)=\mp(1,\pm i)/\sqrt{2}, \nonumber\\
        &&\varepsilon^\mu(0)={1\over M_0}({-M_0^2+P_\bot^2\over
                P^+},P^+,P_\bot),
\end{eqnarray}
We normalize the meson state as
\begin{equation}
        \langle M(P',S',S'_z)|M(P,S,S_z)\rangle = 2(2\pi)^3 P^+
        \delta^3(\tilde P'- \tilde P)\delta_{S'S}\delta_{S'_zS_z}~,
\end{equation}
so that
\begin{equation}
        \int {dxd^2k_\bot\over 2(2\pi)^3}~|\phi(x,k_\bot)|^2 = 1.
\end{equation}

In principle, the momentum distribution amplitude
$\phi(x,k_\bot)$ can be obtained by solving the light-front
QCD bound state equation\cite{Zhang,Cheung}.
However, before such first-principle
solutions are available, we would have to be contented with
phenomenological amplitudes.  One example that has been often
used in the literature for heavy mesons is the so-called BSW
amplitude\cite{BSW}, which for a $B(b\bar q)$-meson is given by
\begin{equation}
        \phi_B(x,k_\bot) = {\cal N}_B \sqrt{x(1-x)}
                ~{\rm exp}\left({-k^2_\bot\over2\omega_B^2}\right)
                ~{\rm exp}\left[-{M_B^2\over2\omega_B^2}(x-x_0)^2\right],
                \label{bswamp}
\end{equation}
where ${\cal N}_B$ is the renormalization constant, $x$ is the
longitudinal momentum fraction carried by the light anti-quark,
$x_0=({1\over2}-{m_b^2-m_q^2\over2M_B^2})$, $m_b$=$b$-quark mass,
$m_q$= light-quark mass, and $\omega_B$ is a parameter of order
$\Lambda_{QCD}$.

For the pion, we shall adopt the Gaussian type wave function,
\begin{equation}
        \phi_\pi(x,k_\bot)={\cal N}_\pi \sqrt{{dk_z\over dx}}
        ~{\rm exp}\left(-{\vec k^2\over 2\omega_\pi^2}\right),
        \label{gauss}
\end{equation}
where $\vec k=(k_\bot, k_z)$,
\begin{equation}
x = {E_1+k_z\over E_1 + E_2}, \qquad
1-x = {E_2-k_z \over E_1 + E_2},
\end{equation}
with $E_i = \sqrt{m_i^2 + \vec k^2}$.
We then have
\begin{equation}
M_0=E_1 + E_2,
\end{equation}
\begin{equation}
      k_z = \left(x-{1\over2}\right) M_0
            - {m_1^2-m_2^2 \over 2 M_0}
\end{equation}
and
\begin{equation}
        {{dk_z\over dx}} = {E_1 E_2\over x(1-x)M_0}
\end{equation}
is the Jacobian of transformation from $(x, k_\bot)$ to
$\vec k$.
This wave function has been also used in many other studies
of hadronic transitions.
In particular, with appropriate parameters, it
describes satisfactorily the pion elastic form factor up
to $Q^2\sim 10~{\rm GeV}^2$ \cite{Chung}.

With the light-front wave functions given above, and taking
a Lorentz frame where $P_{B\bot}=P_{\pi\bot}=0$ (i.e., $q_\bot=0$),
we readily obtain
(for $B^0\rightarrow \pi^+l^-\bar\nu$)
\begin{eqnarray}
        \langle\pi(P_\pi)|J^\mu|B_0(P_B)\rangle =\sum_{\lambda's}
                \int && \{d^3p_d\}
                \phi^*_\pi(x',k_\bot) \phi_B(x,k_\bot)\nonumber\\
                && \times R^{00\dagger}_{\lambda_u\lambda_d}(x',k_\bot)
                 ~ \bar u(p_u,\lambda_u)\gamma^\mu u(p_b,\lambda_b)
                R^{00}_{\lambda_b\lambda_d}(x,k_\bot), \label{mel}
\end{eqnarray}
where $x(x')$ is the momentum fraction carried by the
spectator anti-quark ($\bar d$) in the initial(final) state,
such that
\begin{equation}
xP^+_B=x'P^+_\pi;   \label{xpxp}
\end{equation}
the meaning of all the other variables should be self obvious.
The valence-quark part of the reaction mechanism is depicted in
Fig. (1a).
Substituting the covariant form given in Eq. (\ref{covariant})
into Eq. (\ref{mel}), we get
\begin{eqnarray}
        \langle\pi(P_\pi)|J^\mu|B_0(P_B)\rangle =
                \sqrt{P^+_B\over P^+_\pi}
                \int &&{dx d^2k_\bot \over2(2\pi)^3}
                \phi^*_\pi(x',k_\bot) \phi_B(x,k_\bot)
                {-1 \over 2 \tilde M_{0\pi} \tilde M_{0B}
                \sqrt{(1-x')(1-x)} } \nonumber\\
                && \times {\rm Tr}\left[\gamma_5(\not{\! p}_u+m_u)\gamma^\mu
                (\not{\! p}_b+m_b) \gamma_5 (\not{\! p}_d-m_d)\right].
\end{eqnarray}
The trace in the above expression can be easily carried out.
For the ``good" component, $\mu=+$, we get
\begin{eqnarray}
        -{\rm Tr}&&\left[\gamma_5(\not{\! p}_u+m_u)\gamma^+ (\not{\! p}_b
                +m_b) \gamma_5 (\not{\! p}_d-m_d)\right] \nonumber\\
                && =
            4 \left[{m_q^2+k_\bot^2\over x}+m_q(m_b-m_q)\right] P_\pi^+,
\end{eqnarray}
where we have set $m_u=m_d=m_q$.
From Eq. (\ref{xpxp}), $x$ and $x'$ are related by
$x=R_\pm x'$, with
\begin{equation}
        R_\pm \equiv
                {P_\pi^+ \over P_B^+} = {1 \over 2M_B^2} \Big[ M_B^2
                +M_\pi^2 -q^2 \pm \sqrt{(M_B^2 +M_\pi^2 -q^2)^2
                - 4 M_B^2 M_\pi^2}~\Big].   \label{rq2}
\end{equation}
$R_\pm$ correspond to the pion recoiling in the positive
and negative z-direction respectively relative to the B meson.
The physical kinematic
range $q^2: 0\rightarrow (M_B-M_\pi)^2$ corresponds to
 \begin{eqnarray}
 R_+ && : 1 \rightarrow M_\pi / M_B, \nonumber\\
 R_- && : (M_\pi/M_B)^2 \rightarrow M_\pi / M_B.
 \label{R+-}
 \end{eqnarray}

As pointed out in Ref. \cite{Cheung},
the choice of the positive z-axis is immaterial,
and the matrix elements calculated
in both reference frames (call them the ``+" and ``$-$" frame)
should produce the same form factors $f_\pm(q^2)$
as defined in Eq. (\ref{mele}).
In the ``+" frame, we write
\begin{equation}
f_+^v(q^2)(P_B^+ + P_\pi^+) + f_-^v(q^2)(P_B^+ - P_\pi^+)
= F(R_+) P_\pi^+,
\end{equation}
or equivalently,
\begin{equation}
f_+^v(q^2)(1 + R_+) + f_-^v(q^2)(1 - R_+)
=F(R_+) R_+; \label{ff+}
\end{equation}
similarly, in the ``$-$" frame,
\begin{equation}
f_+^v(q^2)(1 + R_-) + f_-^v(q^2)(1 - R_-)
=F(R_-) R_-, \label{ff-}
\end{equation}
where
\begin{eqnarray}
                F(R_\pm) = && {\sqrt{R_\pm}\over(2\pi)^3}
                \int_0^{1} dx' \int d^2k_\bot
                \phi^*_\pi(x', k_\bot) \phi_B(R_\pm x', k_\bot)\nonumber\\
                && \times {1\over {\tilde M_{0\pi}} {\tilde M_{0B}}
                \sqrt{(1-x')(1-R_\pm x')}} \left[
                {m_q^2+k_\bot^2\over R_\pm x'}+m_q(m_b-m_q)\right].
               \end{eqnarray}
Solving for $f^v_\pm(q^2)$ from Eqs. (\ref{ff+}, \ref{ff-}),
we finally arrive at
\begin{equation}
f^v_+(q^2)={ F(R_+)R_+(1-R_-) - F(R_-)R_-(1-R_+)
\over 2(R_+ - R_-)}\label{fv+} \\
\end{equation}
and
\begin{equation}
f^v_-(q^2)={-F(R_+)R_+(1+R_-) + F(R_-)R_-(1+R_+)
\over 2(R_+ - R_-)}.\label{fv-}
\end{equation}

These are the valence-configuration contributions
to the $B \rightarrow \pi l {\nu}$
decay form factors, valid in the entire range of
momentum transfer $q^2 = [ 0 ,(M_B - M_\pi)^2 ]$.

\subsection{Higher-Fock-State Contribution}

It is often not adequate to describe the internal structure of a hadron
solely by its valence configuration.
As we have discussed,
for time-like momentum transfers, one must also consider the effects
of higher Fock-states corresponding to configurations containing
quark-antiquark pairs in addition to the valence particles.
Such contributions are shown in Fig. (1b).
Unfortunately these higher-Fork-state wave functions are not available,
and we shall estimate their contributions in an effective mesonic picture.
In the case of the $B$-meson,
it is expected that the effective higher-Fock-state configuration
$|B^* \pi\rangle$ is the most important if the relative momentum
is small.
The reason is as follows.
The $B$ and $B^*$ masses are almost degenerate
due to heavy-quark symmetry, and also the pion mass
is small. Consequently, when the relative momentum is small,
the $|B^*\pi\rangle$ configuration is
close to the energy shell (i.e., the energy denominator is small),
and is thus enhanced.  This can be readily seen in Eq.(\ref{wf}),
where the interaction Hamiltonian $H_I$ describing the
$BB^*\pi$ coupling is given in chiral perturbative theory by\cite{Yan}
\begin{equation}
        H_I = -{g\over f_\pi}\sqrt{M_B M_{B^*}}~
                \int dx^- d^2 x_\bot B^{*\dagger}_\mu
                (i\partial^\mu\pi^a)\tau^a B,
                \label{HI}
\end{equation}
with $f_\pi$ = 93 MeV being the pion decay constant.

The contribution to the decay matrix element
from the $|B^*\pi\rangle$ configuration [see Fig. (1c)]
is given by
\begin{equation}
        \langle\pi(P_\pi)|J^\mu|B(P_B)\rangle_{B^*} =
                \int [d^3q] \langle 0| J^\mu |B^*(q)\rangle
                {\langle B^*(q) \pi(P_\pi)|H_I|B_0 (P_B) \rangle
                \over P^-_B-q^- -P^-_\pi} .
\end{equation}
From
\begin{eqnarray}
         \langle 0| J^\mu |B^*(q)\rangle &=&  M_{B^*} f_{B^*}
                \varepsilon^{\mu},  \\
         \langle B^*(q) \pi^-(P_\pi)|H_I|B_0 (P_B) \rangle &=&
                2(2\pi)^3 \delta^3 (\tilde{P}_B - \tilde q -
                \tilde{P}_\pi) {\sqrt{2} g \over f_\pi}
                \sqrt{M_B M_{B^*}} ~ \varepsilon^* \cdot P_\pi  ,
\end{eqnarray}
and
\begin{equation}
        \varepsilon^\mu \varepsilon^{*\nu} = - g^{\mu \nu}
                + {P_{B^*}^\mu P_{B^*}^\nu \over M_{B^*}^2},
\end{equation}
we readily obtain
\begin{equation}
        \langle\pi^-(P_\pi)|J^\mu|B(P_B)\rangle_{B^*} =
                \sqrt{2} g {f_{B^*} \over
                f_\pi} {P_{B^*} \cdot P_\pi P_B^\mu
                - (M_{B^*}^2 + P_{B^*} \cdot P_\pi) P_\pi^\mu
                \over P_{B^*}^+ (P_B^- - P_{B^*}^-
                - P_\pi^-) }\sqrt{M_B \over M_{B^*}}.
\end{equation}
Therefore,
\begin{eqnarray}
        && f^{B^*}_+ (q^2) = g {f_{B^*} \over \sqrt{2} f_\pi}
                {-M_{B^*}^2 \over P_{B^*}^+ (P_B^- - P_{B^*}^-
                - P_\pi^- )} \sqrt{M_B \over M_{B^*}}
                \label{fb+} \\
        && f^{B^*}_- (q^2) = g {f_{B^*} \over \sqrt{2} f_\pi}
                {M_{B^*}^2 + 2P_{B^*} \cdot P_\pi \over P_{B^*}^+
                (P_B^- - P_{B^*}^- - P_\pi^-) } \sqrt{M_B \over M_{B^*}}
                \label{fb-}
\end{eqnarray}
where $P^+_{B^*} = P_B^+ - P_\pi^+ > 0 $ and
$P_{B^*}^- = { M_{B^*}^2 + (P_{B\bot}- P_{\pi\bot})^2
\over P_B^+ - P_\pi^+}$.
It is easy to see that $f_\pm^{B^*}$ are functions of $q^2$ because
\begin{equation}
        {1 \over q^2 - M_{B^*}^2} = {1 \over P^+_{B^*}}
                {1 \over P_B^- -P_{B^*}^- - P_\pi^-} ~~~~~
                (q^+=P^+_{B^*}). \\
\end{equation}

However the above results are not quite complete, because
chiral perturbation theory is a low-energy effective theory,
such that the chiral $BB^*\pi$-vertex given in Eq. (\ref{HI})
is valid only for soft pions.  A suppression factor
is generally expected when the pion momentum increases.
This can also be understood in the quark picture by the following reasoning.
The higher Fock-state $|B^*\pi\rangle$
arises from quark-pair creation
which is a predominently soft process.
First of all, there is not much probability for producing
hard $u\overline{u}$ pairs. Moreover, once produced,
a hard $u$ has little chance of forming a pion with a slow $\bar u$,
likewise a hard $\overline{u}$ is not likely to
form a $B^*$-meson with a slow $b$ quark.
Therefore configurations with a high-momentum pion in
$|B^*\pi\rangle$ is expected to be suppressed.
This physical requirement can be implemented phenomenologically by
introducing a damping form factor, ${\cal F}(q^2)$,
to the chiral $BB^*\pi$-vertex, such that,
\begin{equation}
        g\rightarrow g {\cal F}(q^2). \label{cconst}
\end{equation}
We shall take
\begin{equation}
        {\cal F}(q^2)
        = \exp\Bigg(-{v_B \cdot (p_\pi-M_\pi)\over\Lambda}\Bigg),
                    \label{ffactor}
\end{equation}
where $v_B=P_B/M_B$ is the 4-velocity of the $B$-meson,
and $\Lambda$ is the cutoff momentum which is expected to be
of the order of the chiral symmetry breaking scale
$\Lambda_\chi \simeq 1$ GeV.
The form of the suppression factor is similar to that proposed
by Wolfenstein \cite{Wol} in the soft-pion region, and is normalized to
unity at the zero-recoil point where $q^2=(M_B-M_\pi)^2$.
The substitution indicated in Eq. (\ref{cconst}) is to be understood
for all of the equations derived in this Subsection.

\subsection{Wave Function Renormalization}

Finally we calculate the wave function renormalization constant $Z_2$
which is required for consistency,
as indicated in Eq. (\ref{wf}).  The result is
\begin{eqnarray}
        Z_2^{-1} &=& 1 + {1 \over 2(2\pi)^3 P_B^+} \int [d^3k]
                [d^3q]~{|\langle B^*(q) \pi(k)|H_I|B_0 (P_B)
        \rangle|^2 \over (P^-_B-q^- -k^-)^2} \nonumber \\
                &=& 1 + {3 g^2 \over f_\pi^2} {M_BM_{B^*}
        \over P_B^+} \int [d^3q]~{1 \over k^+}~{[\varepsilon
                \cdot k]^2 \over [P_B^- - q^- -
                k^-]^2} {\cal F}^2[(P_B-k)^2], \label{z2}
\end{eqnarray}
where $q$ and $k$ are on the mass-shells, $k^+=P_B^+-q^+$,
$k_\bot=-q_\bot$,
and we have included the suppression form factor
as given in Eqs. (\ref{cconst},\ref{ffactor}).
%Note that the same result can also be obtain by calculating the
%$B$-meson self-energy Feynman diagram with a pion loop.

Combining Eqs.(\ref{fv+}), (\ref{fv-}), (\ref{fb+}),
(\ref{fb-}) and (\ref{z2}) together, we obtain a relatively
more consistent treatment for $B \rightarrow \pi l {\nu}$ decay
form factors in comparison to the previous studies.

%%%%%%%%%%%%%%%%%%%%%%%%%%%%%%%%%%%%%%%%%%%%%%%%%%%%%%%%%%%%%%%%%%%%%%%%%%%%%%%

\section{Numerical Results}

In this section, we present the numerical results for
$B \rightarrow \pi l {\nu}$ decay form factors.  We first present
contributions from the valence-quark configuration.
The parameters in the light-front wave functions
are fixed by fitting to other hadronic properties.
For the pion wave function, the parameters $\omega_\pi$
and $m_q(=m_{u,d})$ are fitted to the pion decay constant $f_\pi$
and also the pion elastic form factor for momentum
transfer $Q^2=0 \sim 10$ GeV$^2$ \cite{Chung}:
\begin{equation}
        \omega_\pi = 0.29 ~{\rm GeV}, ~~~
         m_q = 0.30 ~{\rm GeV}.
%         ~~~ M_\pi = .140 ~{\rm GeV}.
\end{equation}
For the light-front $B$-meson wave function, we take
\begin{equation}
        \omega_B = 0.57 ~{\rm GeV}, ~~~
        m_b = 4.8 ~{\rm GeV},
%~~~ M_B =5.278 ~{\rm GeV}.
\end{equation}
which were determined from the $B$ decay constant $f_B=0.187$ GeV, and
other decay data\cite{BSW,Odon2}.  With these parameters,
the decay form factor $f^v_+$ due to the valence configuration
is presented in Fig. (2).
At the maximum recoil point ($q^2=0$),
\begin{equation}
        f^v_+ (0) = 0.24,              \label{q0}
\end{equation}
which is consistent with results from most other calculations
\cite{BSW,ISGW,Odon,Faus,Ball,Che,Bel1,Col,Ali,HYC,Neu,LY}.
A slightly different set of parameters for the pion wave function
\begin{equation}
        \omega_\pi = 0.33 ~{\rm GeV}, ~~~
         m_q = 0.25 ~{\rm GeV}
\end{equation}
also fits the pion data quite well.
The result is not qualitatively changed
by using this set of numbers, as can be seen also in Fig. (2).
The value of the form factor at $q^2=0$ is however increased by
20\% to
\begin{equation}
f^v_+(0) = 0.29.                \label{q00}
\end{equation}

It is a common practice in the literature to extrapolate $f_+(q^2)$
away from the $q^2=0$ point by a monopole $q^2$-dependence,
even though such a $q^2$-dependence is expected to be reasonable only
near the zero-recoil point [$q^2 = (M_B-M_\pi)^2$].
Our result does not confirm such a $q^2$-dependence.
Instead, as shown in Fig. (4), for a wide range of momentum transfer,
$q^2 = 0\sim 18~{\rm GeV}^2$, the valence contribution
agrees rather well with the following formula:
\begin{equation}
        f_+^{pole}(q^2) = {f_+(0) \over (1 - q^2/M_{pole}^2)^{\alpha} } ,
                \label{fit}
\end{equation}
with $\alpha=1.6$ and $M_{pole}=5.32$ GeV.

From Fig. (2), we see that, near the zero-recoil point,
the valence-quark prediction decreases as $q^2$ increases,
and no longer bears any resemblence to Eq. (\ref{fit}).
The dipping of the valence-quark contribution toward the zero-recoil
point can be understood as follows.
Recall that the decay amplitude involves an overlapping integral of the
wave functions of the initial and final mesons.
If both mesons were heavy, then it is obvious that,
by heavy-quark symmetry, maximum overlapping must occur
at the zero-recoil point.
However, in the present situation,
a heavy ($B$) and a light meson ($\pi$) are involved, and their internal
momentum distributions peak at different values of $x$.
Specifically, $\phi_B(x,k_\bot)$ has a narrow peak near $x=0$,
whereas $\phi_\pi(x,k_\bot)$ peaks with a much larger width at $x=1/2$.
Consequently maximum overlapping of the wave functions actually
occurs somewhat away from the zero-recoil kinematics.

%Note that there is no compelling physical reason for
%the monopole behavior of the valence-quark
%prediction in the region $q^2=0\sim 20~{\rm GeV}^2$,
%our result comes from overlapping the quark-momentum distributions
%in $B$ and $\pi$.
%It is interesting to note that a similiar result
%has also been obtained
%in a study using the QCD sum rule approach\cite{Ball}.
%However, for us, this is not the whole story, since we have not yet
%included the higher-Fock-state contribution.

To test the sensitivity to the $B$-meson wave function used,
we plot in Fig. (3) the valence contribution,
using a Gaussian wave function of the same form as shown
in Eq. (\ref{gauss}), with $\omega_B=0.73$ GeV.
We find that the result is not changed significantly
except in the zero-recoil region, $q^2 > 20$ GeV$^2$.  As we shall see,
in the zero-recoil region, the $B^*$-pole contribution dominates, and
the valence contribution is relatively not important.

Next we consider the
$B^*$-pole (or higher-Fock-state) contribution
given by Eqs. (\ref{fb+}, \ref{fb-}).
We take
\begin{eqnarray}
        && f_B =0.187~ {\rm GeV}, ~~~ f_{B^*} /f_B = 1.1,
         ~~~ g = 0.48,
         ~~~\Lambda=1.0~{\rm GeV},
\label{numval}
\end{eqnarray}
where the values taken for $f_B$ and the ratio $f_B/f_{B^*}$
are consistent with QCD sum rule and lattice QCD
results\cite{HYC,Neubert2,Abada}.
However the magnitude of $g$ is less certain;
it is not directly measurable, since $B^*\rightarrow B\pi$
is kinematically forbidden.
In the infinite-heavy-quark-mass limit,
heavy-quark symmetry predicts that $g$ is independent of heavy quark flavor,
so that $BB^*\pi$ is equal to $DD^*\pi$ which is experimentally measurable.
However,they have different $1/m_Q$ corrections.
From the upper limit of the total width of
$D^{*+} (<131$ MeV)\cite{ACCMOR}, we know that,
$g$ should be less than 0.7.
From non-relativistic quark-model constrained by
the axial coupling constant $g_A=1.25$ of the nucleon,
one gets\cite{Yan,Cheng2} $g=0.75$
in the symmetry limit,
which some what overshoots the experimental upper limit.
QCD-sum-rule calculations\cite{Aliev,Colangelo,Belyaev2}
usually obtain smaller values, namely,
$g\simeq0.2\sim 0.4$.
Results from various other approaches tend to fall
between these two limits \cite{Odon2,Cho,Amundson,Bardeen,Eichten}.
In Eq. (\ref{numval}), we have simply taken
the average of the non-relativistic
quark model prediction and the lowest of the QCD sum-rule results.
The cutoff momentum $\Lambda$ in Eq. (\ref{cconst}) is also not
well known, and we have assumed $\Lambda$ to be equal to
the chiral symmetry breaking scale $\Lambda_\chi\simeq 1$ GeV.
Sensitivity of the results to the values of $g$ and $\Lambda$
will be shown below.

Finally, we need to compute the renormalization constant $Z_2$
given by Eq. (\ref{z2}).
The result is $\sqrt{Z_2}$ = 0.85, 0.93, 0.98,
for $g$ = 0.75, 0.48, 0.20 respectively.
Hence, depending on the coupling constant $g$, the effect of wave
function renormalization can be quite significant (2-15\%).

The $B^*$-pole contribution is plotted in Fig. (5a)
with $\Lambda = 1.0$ GeV and three values of $g$ as indicated.
The combined valence and $B^*$-pole contribution
is plotted in Fig. (5b).
We see that the $B^*$-pole contribution dominates
in the zero-recoil region,
$q^2\simeq 25$ GeV$^2$, and decreases rapidly as $q^2$ decreases.
With the $B^*$-contribution included, we find $f_+(0)=0.26, 0.28, 0.29$
for $g=0.20, 0.48, 0.75$ respectively; therefore the effect of the $B^*$
contribution at $q^2=0$ is within the theoretical error caused by the 
uncertainty in the pion wave function [see Eqs. (\ref{q0}, \ref{q00})].
For $g=0.48$, the theoretical result can be approximately described
by Eq. (\ref{fit}), with $\alpha=2.0$ and $M_{pole}=6.0$ GeV
[see Fig. (6a)].
The $g=0.2$ curve also shows approximately
a dipole behavior ($\alpha=2.0$),
with a slightly different pole mass $M_{pole}=5.8$ GeV [see Fig. (6b)].
Hence our results indicate that $f_+(q^2)$ does not follow
a simple monopole $q^2$-dependence (i.e. $\alpha=1.0$).

In order to study the sensitivity of our results
to the cut-off parameter $\Lambda$,
we have plotted in Fig. (7) the same curves as in Fig. (5b),
but with $\Lambda$=1.5 GeV.
From Eqs. (\ref{fb+},\ref{cconst},\ref{ffactor}), it is easily seen that
the $B^*$-pole contribution $f_+^{B^*}$ increases with $\Lambda$;
this effect is however counter balanced by a smaller $Z_2$.
Since we have taken into account
the effect of wave function renormalization, our results are
relatively insensitive to the precise value of $\Lambda$ used.
It thus seems that a measurement of $f_+(q^2)$ near zero-recoil
would give a rather good estimate
of the $BB^*\pi$ coupling constant $g$,
which is not obtainable from direct strong decays.

\section{Summary}

The exclusive $B \rightarrow \pi l {\nu}$ decay is studied in this paper.
We have calculated two different contributions to the decay process:
The valence-quark contribution is calculated using light-front
wave functions for mesons, while the $B^*$-pole contribution
is evaluated with the help of chiral perturbation theory.
In contrast to previous works using relativistic wave functions,
the decay form factors have been calculated directly in the
entire range of momentum transfer,
so that extrapolation from $q^2=0$ is no longer required.
Furthermore, in a more unified approach,
the $B^*$-pole contribution is calculated from
the higher-Fock-state configuration $|B^*\pi\rangle$ of
the $B$-meson wave function.
The mixing of different configurations
requires a consistent renormalization of the $B$-meson wave function,
which is a 7\% effect for $g$=0.48; 2\% if $g$=0.2.
We find that the form factor $f_+(q^2)$ does not follow a monopole
$q^2$-dependence, as is customarily assumed in the literature.
Instead, our result is closer to a dipole $q^2$-dependence.
At the maximum recoil point, we find that $f_+(0)=0.24-0.29$, which is 
consistent with most other calculations.
Finally we observe that, since the $B^*$-pole contribution dominates
in the zero-recoil region, a measurement of $f_+(q^2)$ near zero-recoil
would be helpful in determining the value of the chiral $BB^*\pi$
coupling constant $g$.

\acknowledgments
It is a pleasure to thank H.Y. Cheng for helpful discussions.
We would also like to thank S.J. Brodsky for his comment on
light-front calculations with time-like momentum transfers.
One of the authors (CYC) thanks the MIT Department of
Mathematics for hospitality, where part of this work was completed.
This work was supported in part by the National Science Council of
the Republic of China under Grant Nos. NSC84-2112-M-001-036,
NSC85-2112-M-001-023, and NSC85-2816-M-001-001L.
%
%
%%%%% References %%%%%%%%%%%%%%%%%%%%%%%%%%%%%%%%%%%%%%%%%%%%%%%%%%%%%%%%%%%%%
%
%
\newcommand{\bi}{\bibitem}
%

%
%
%%%%% Figure Captions %%%%%%%%%%%%%%%%%%%%%%%%%%%%%%%%%%%%%%%%%%%%%%%%%%%%%%%%
%
\newpage
\parindent=0 cm
\centerline{\bf FIGURE CAPTIONS}
\vskip 0.5 true cm

{\bf Fig. (1a) }  $B\rightarrow \pi l\nu$: Valence-quark contribution.
\vskip 0.25 true cm

{\bf Fig. (1b) }  $B\rightarrow \pi l\nu$: Non-valence contribution.
\vskip 0.25 true cm

{\bf Fig. (1c) }  $B\rightarrow \pi l\nu$: $B^*$-pole contribution.
\vskip 0.25 true cm

{\bf Fig. (2) }  Valence-quark contribution to $f_+(q^2)$.
     Solid line:  $m_q=0.30$ {\rm GeV}, $\omega_\pi=0.29$ GeV;
     dashed line: $m_q=0.25$ {\rm GeV}, $\omega_\pi=0.33$ GeV.
\vskip 0.25 true cm

{\bf Fig. (3) }  Solid line: Same as that in Fig. (2);
        dashed line: Gaussian wave function used for $B$, with
        $\omega_B=0.73$ GeV.
\vskip 0.25 true cm

{\bf Fig. (4) }  Solid line: Same as that in Fig. (2);
        dashed line: Eq. (\ref{fit}), with $\alpha=1.6$ and
        $M_{pole}=5.32$ GeV.
\vskip 0.25 true cm

{\bf Fig. (5a) }  Solid lines: $B^*$-pole contribution with $\Lambda=1$ GeV;
dashed line:  valence-quark contribution [same as solid line in Fig. (2)].
\vskip 0.25 true cm

{\bf Fig. (5b) }  Combined valence-quark and $B^*$-pole contribution with
          $\Lambda=1$ GeV.
\vskip 0.25 true cm

{\bf Fig. (6a) }  Solid line: same as the $g=0.48$ line in Fig. (5b);
  dashed line: Eq. (\ref{fit}) with $\alpha=2.0$ and $M_{pole}=6.0$ GeV;
  dash-dotted line: Eq. (\ref{fit}) with $\alpha=1.0$ and $M_{pole}=5.32$ GeV.
\vskip 0.25 true cm

{\bf Fig. (6b) }  Solid line: $g=0.20$ line from Fig. (5b);
  dashed line: Eq. (\ref{fit}) with $\alpha=2.0$ and $M_{pole}=5.8$ GeV;
  dash-dotted line: Eq. (\ref{fit}) with $\alpha=1.0$ and $M_{pole}=5.32$ GeV.
\vskip 0.25 true cm

{\bf Fig. (7) }  Same as Fig. (5b) except that $\Lambda=1.50$ GeV
\vskip 0.25 true cm

%{\bf Fig. (8) }  Same as Fig. (5b) except that $\Lambda=0.75$ GeV
%\vskip 0.25 true cm

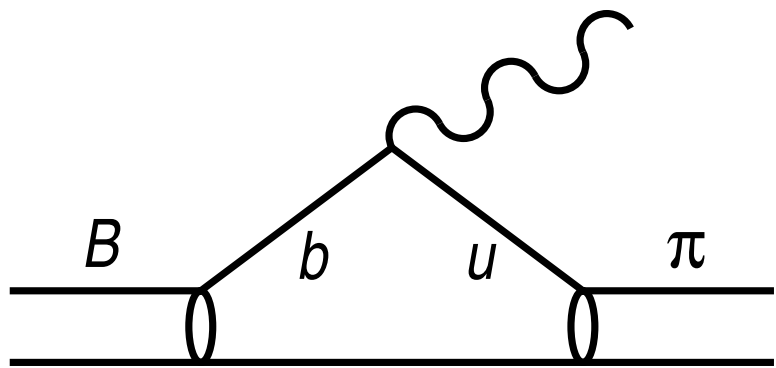
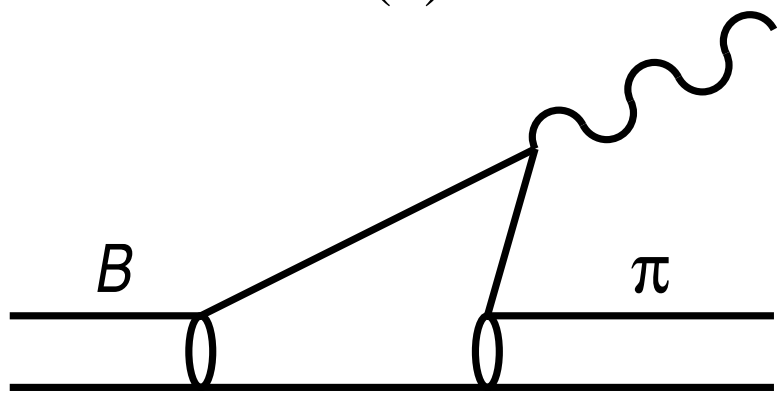
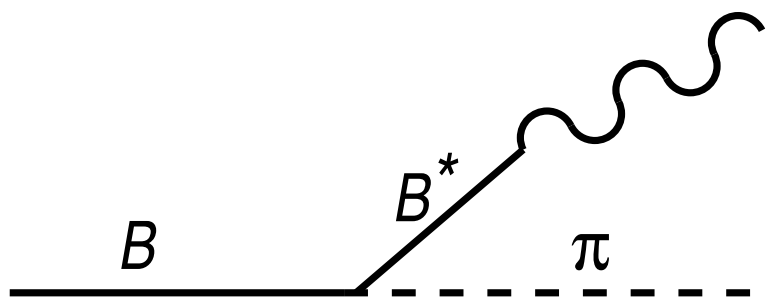
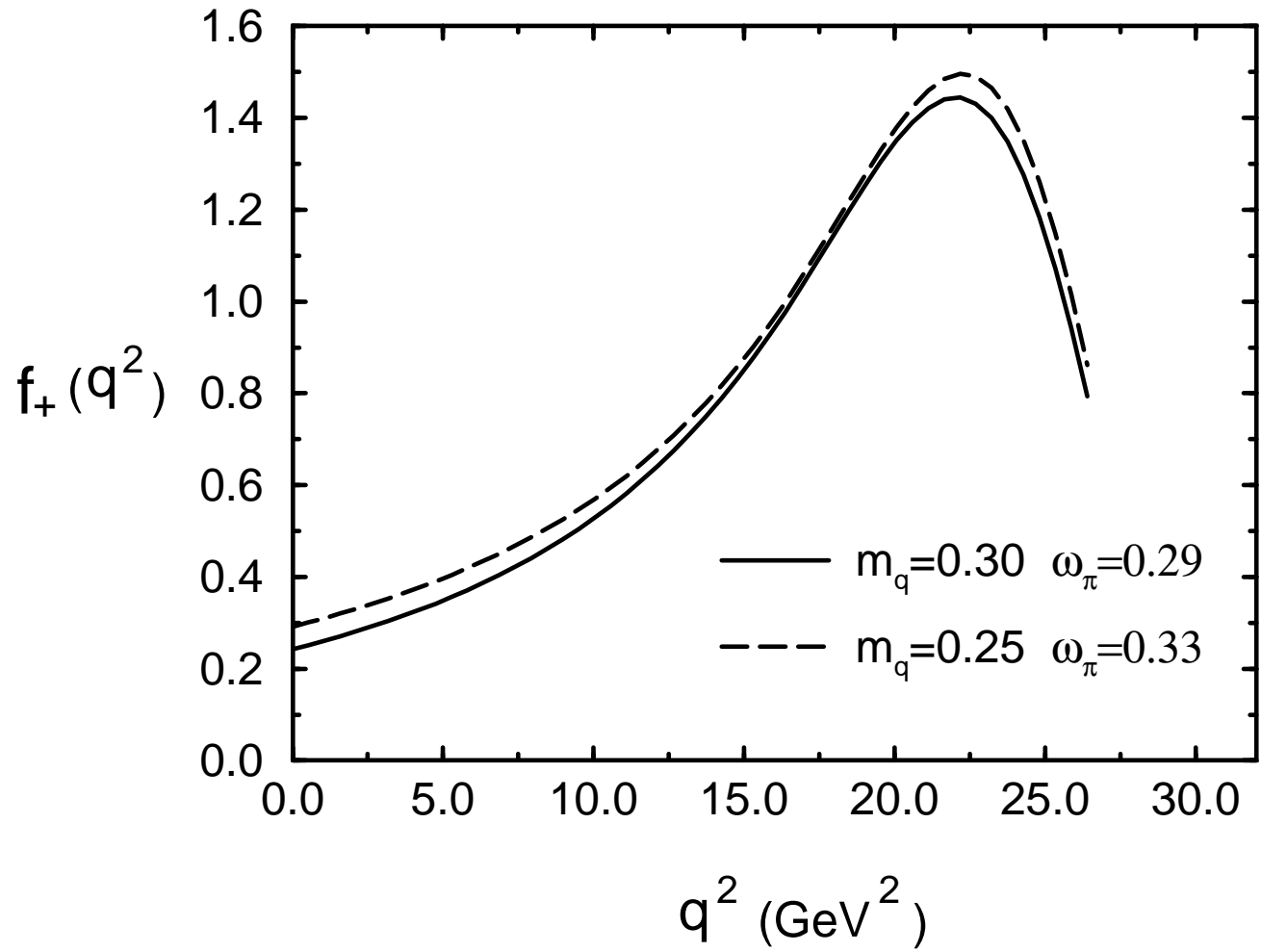
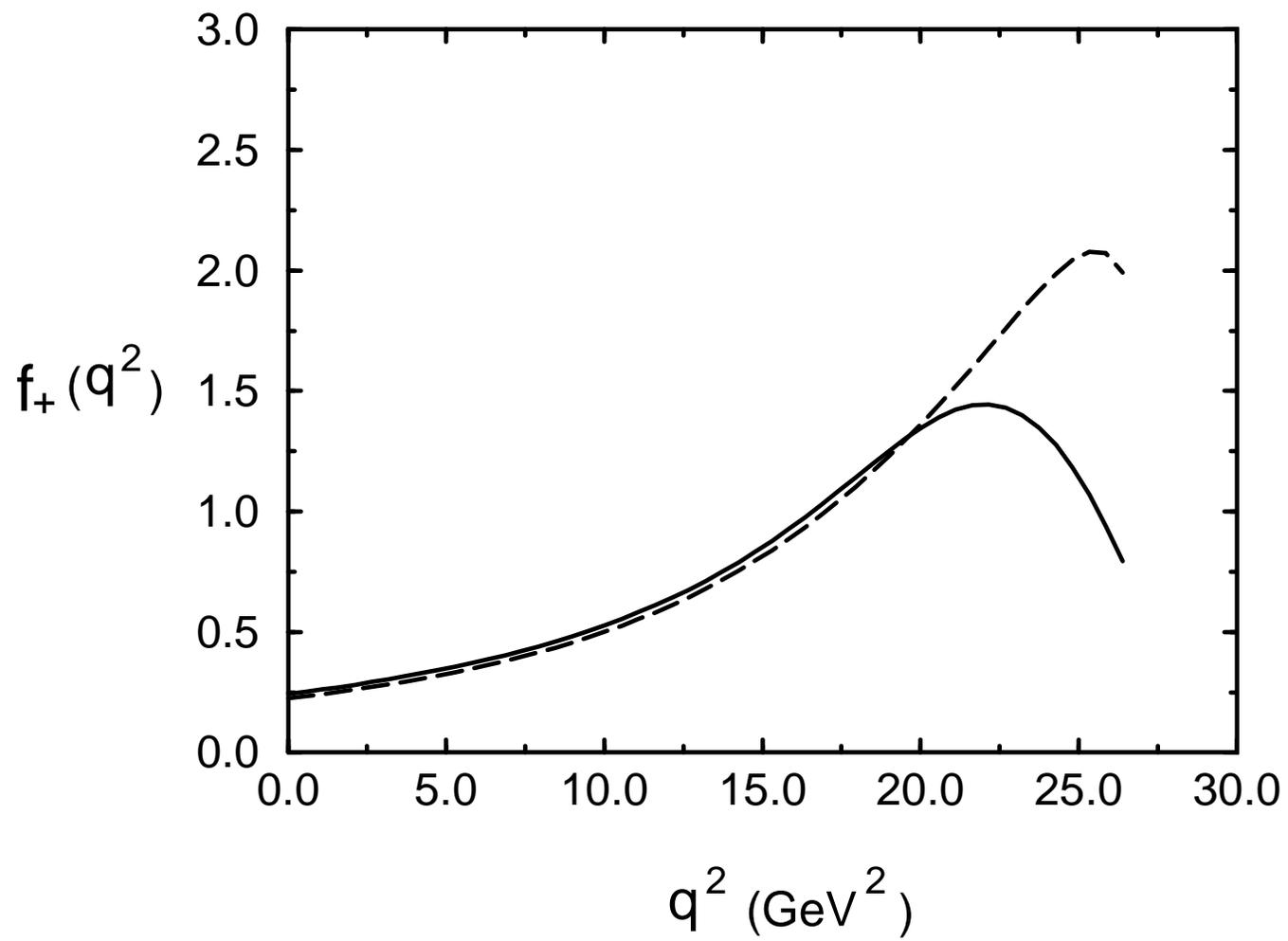
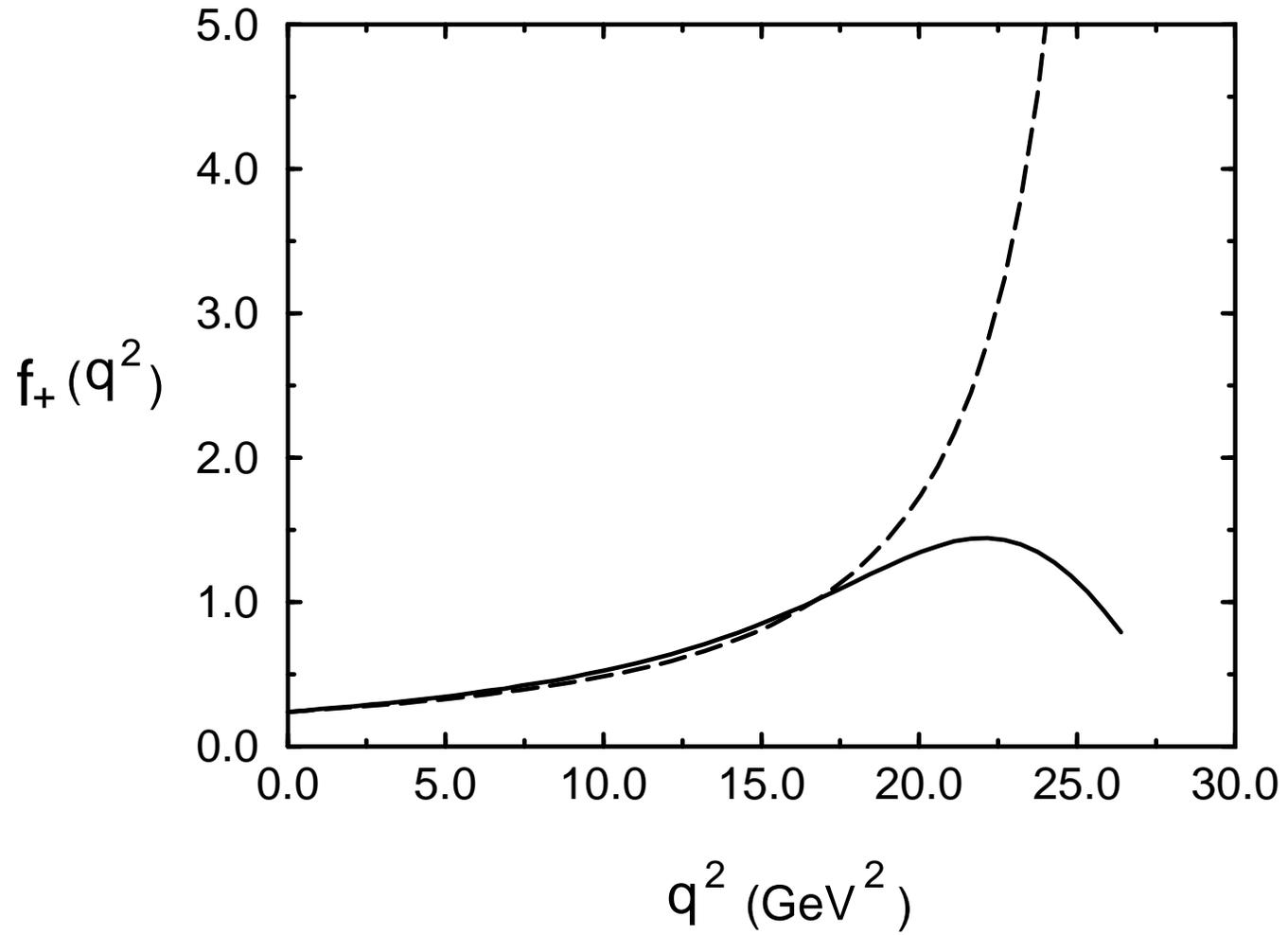
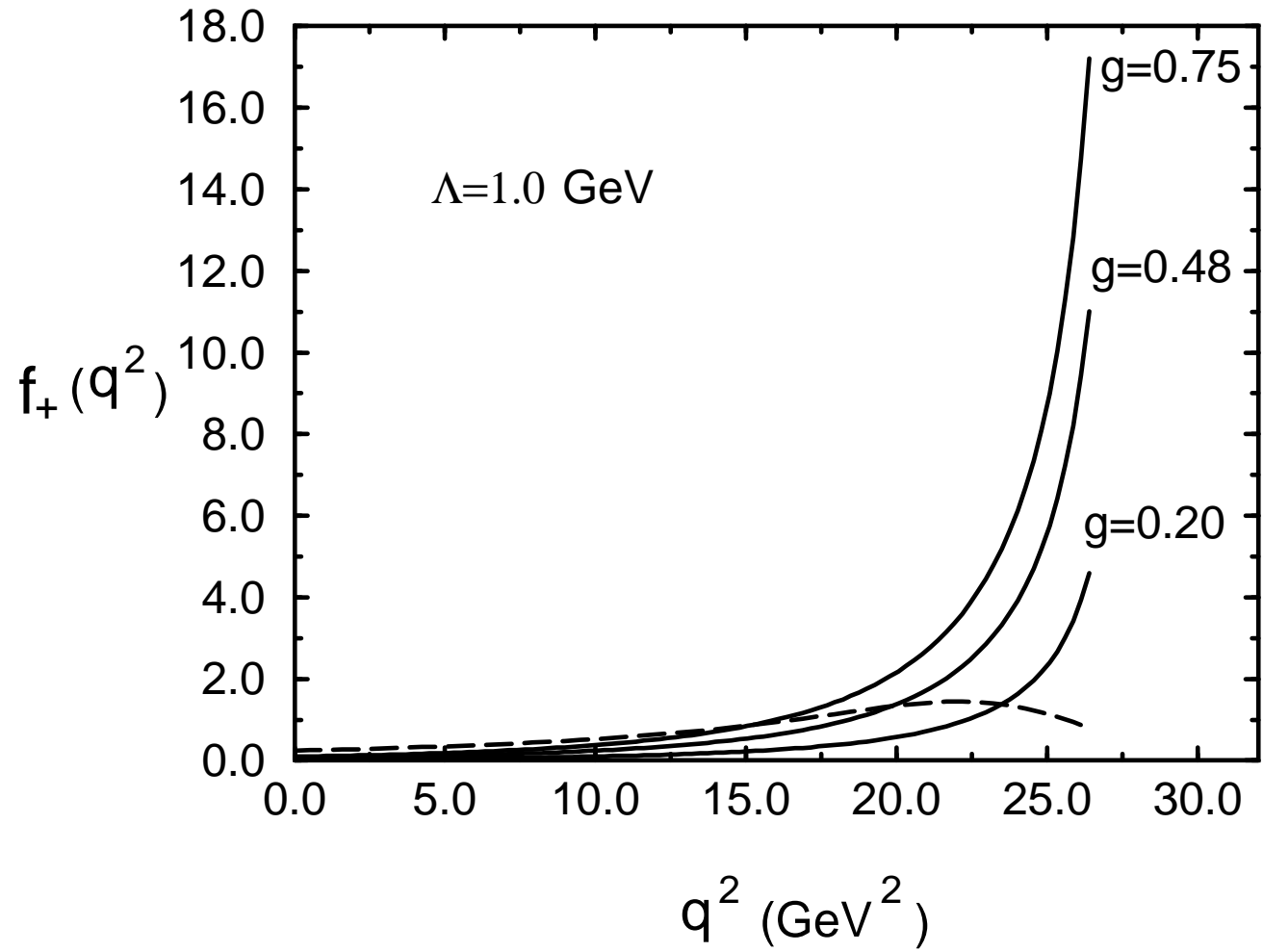
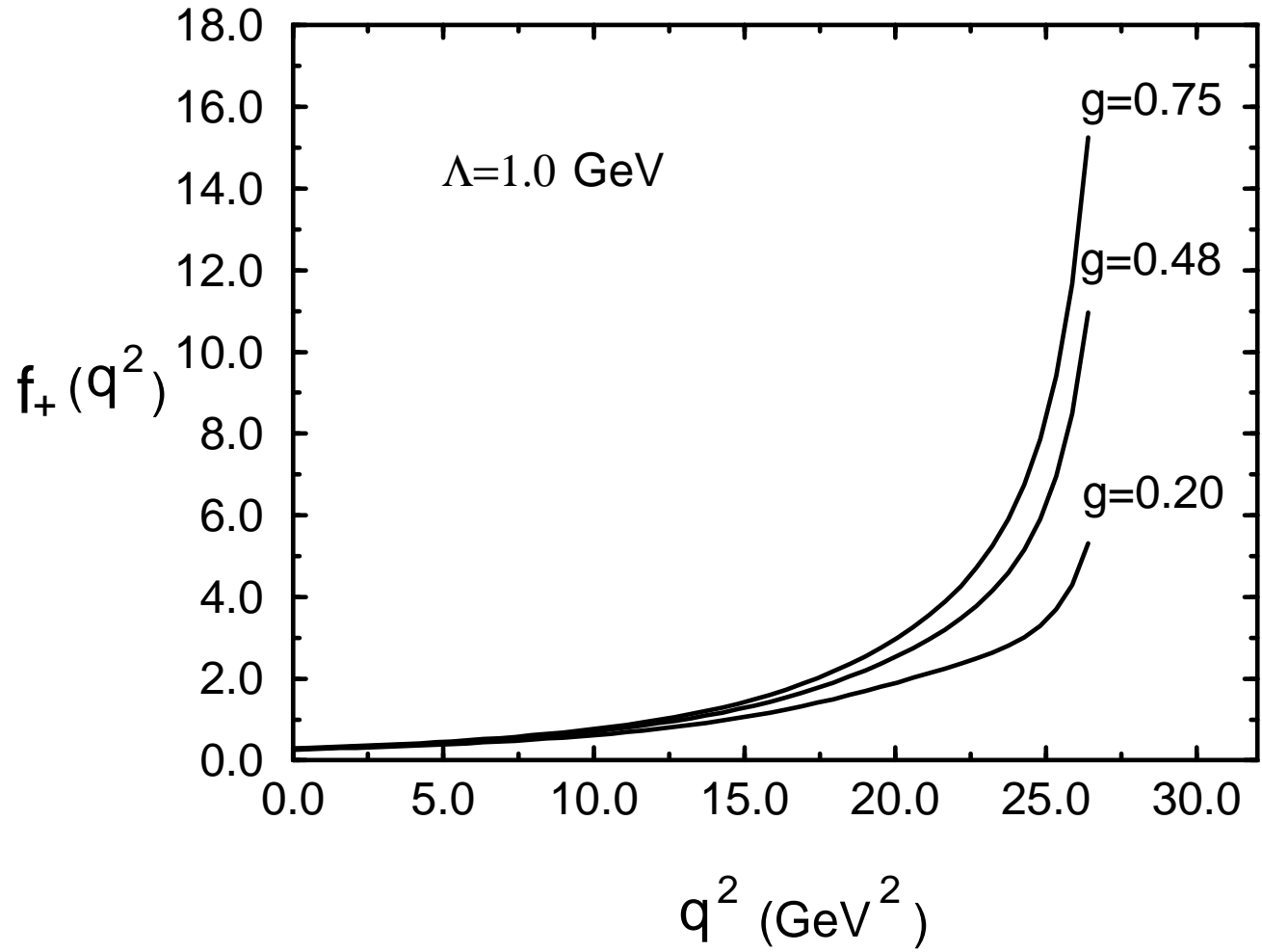
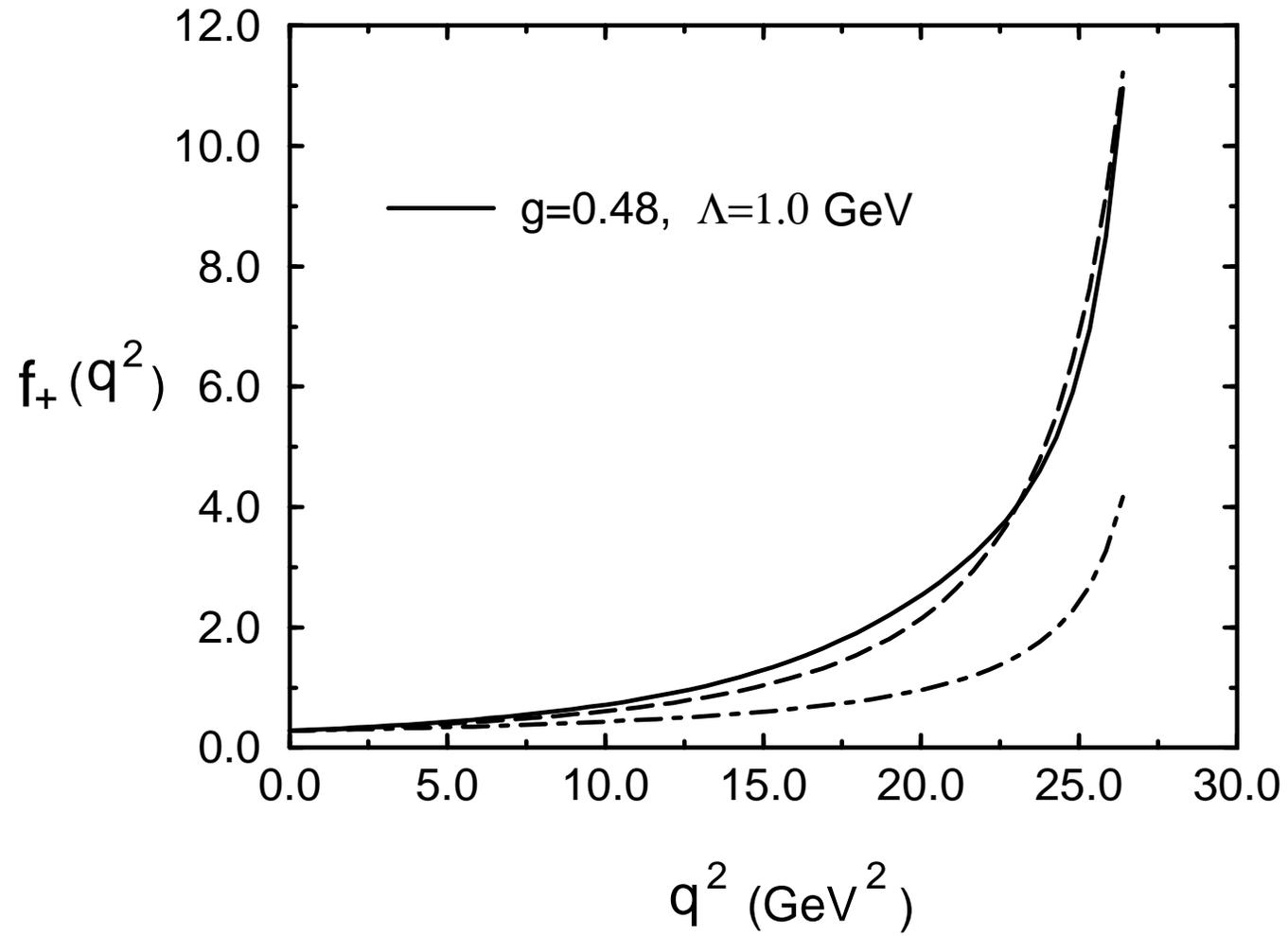
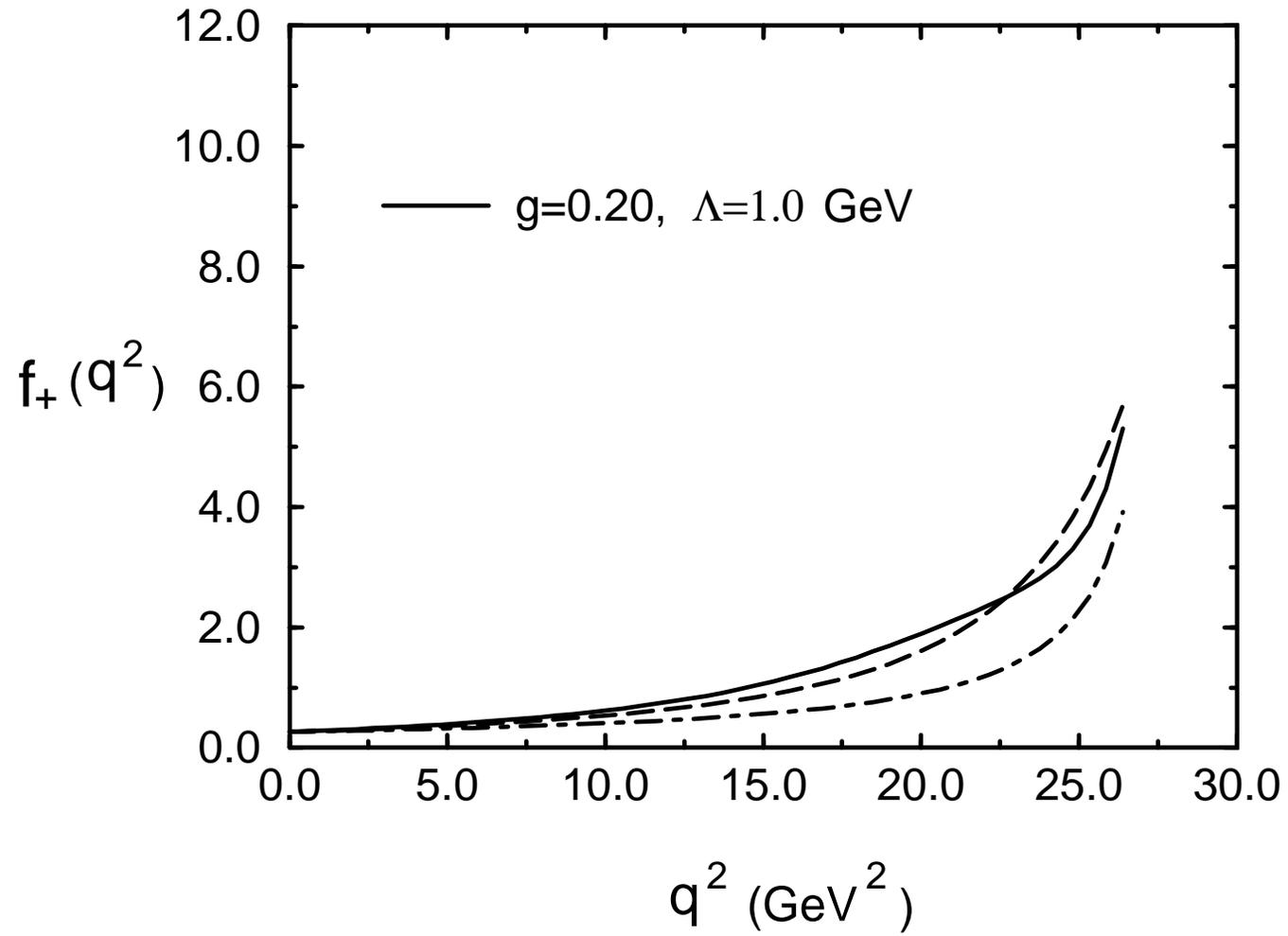
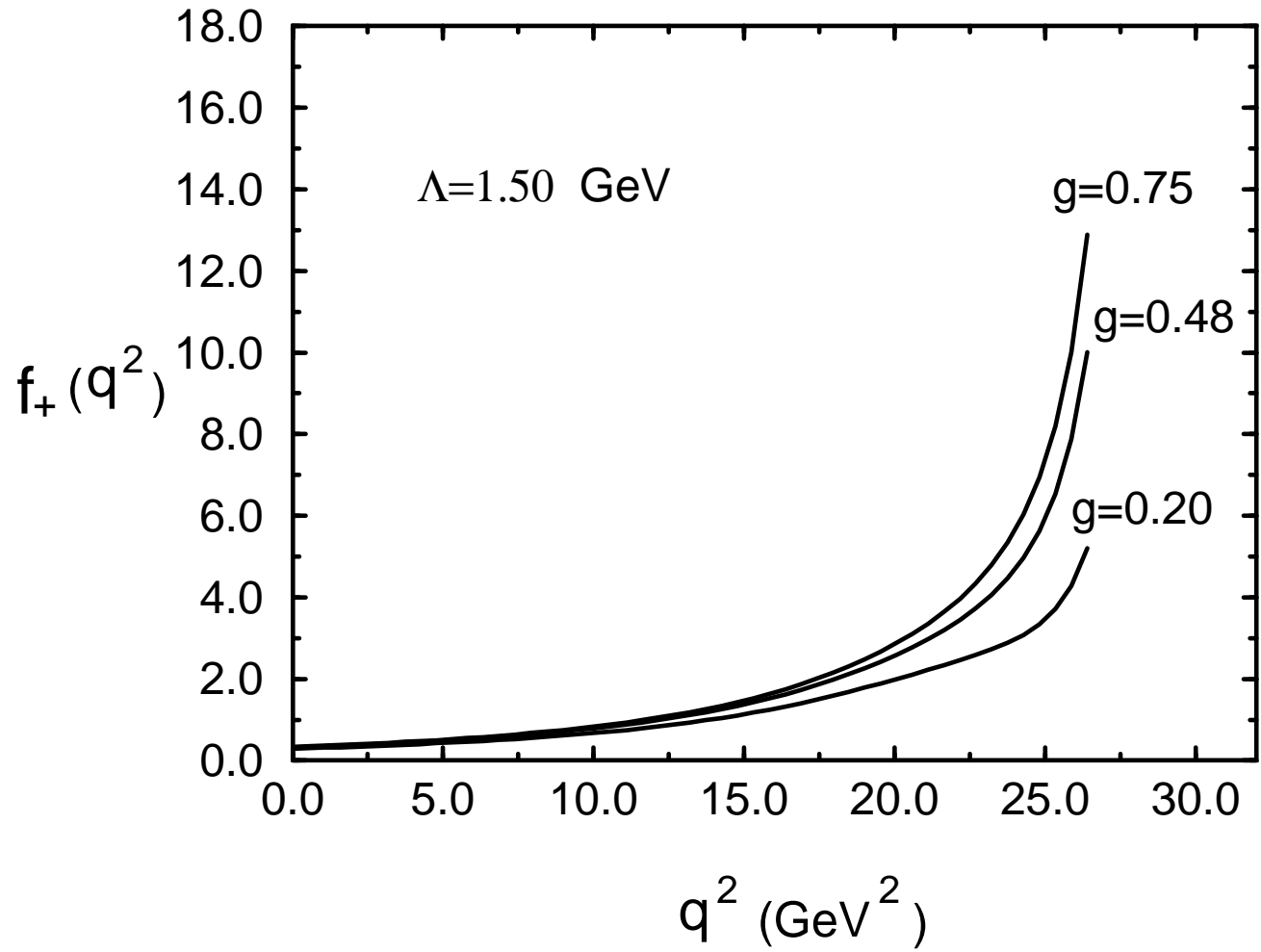
\end{document}